\documentclass[12pt]{elsarticle}
\usepackage{graphicx}
\usepackage{amssymb}

\def\l{\hbox{$l$-index}\/}
\def\h{\hbox{$h$-index}\/}
\def\etal{{\em et~al.}}

\def\USP{Universidade de S\~ao Paulo}
\def\FFCLRP{Faculdade de Filosofia, Ci\^encias e Letras de Ribeir\~ao Preto}

\journal{Physica A}

\begin{document}

\begin{frontmatter}

\title{Lobby index as a network centrality measure}

\author[ifsc]{M.G.~Campiteli}
\author[dcm]{A.J.~Holanda\corref{c2}}
\ead{aholanda@usp.br}
\author[df]{L.D.H.~Soares}
\author[df]{P.R.C.~Soles}
\author[df]{O.~Kinouchi\corref{c1}}
\ead{osame@usp.br}

\cortext[c1]{Principal corresponding author. Phone number: $+$55 16 3602-3779}
\cortext[c2]{Corresponding author}

\address[ifsc]{Instituto de F\'isica de S\~ao Carlos, \USP, Brazil} 

\address[dcm]{Departamento de Computa\c{c}\~ao e Matem\'atica, \FFCLRP, \USP, Brazil}

\address[df]{Departamento de F\'isica, \FFCLRP, \USP, Av.\
  Bandeirantes, 3900, Ribeir\~ao Preto -- SP, 14040-901, Brazil}

\begin{abstract}
  We study the lobby index (\l{} for short) as a local node centrality
  measure for complex networks. The \l{} is compared with degree (a
  local measure), betweenness and Eigenvector centralities (two global
  measures) in the case of a biological network (Yeast interaction
  protein-protein network) and a linguistic network (\textit{Moby
    Thesaurus II}). In both networks, the \l{} has poor correlation
  with betweenness but correlates with degree and Eigenvector
  centralities. Although being local, the \l{} carries more
  information about its neighbors than degree centrality. Also, it
  requires much less time to compute when compared with Eigenvector
  centrality. Results show that the \l{} produces better results than
  degree and Eigenvector centrality for ranking purposes.

\end{abstract}


\end{frontmatter}

\section{Introduction}

The Hirsch index~(\h) has been thoroughly studied for scientometrics
purposes. It has been applied to networks of individual researchers
collaboration~\cite{Hirsch:2005, Hirsch:2007, Batista:2006,
  Bornmann:2007, Alonso:2009}, research groups~\cite{vanRaan:2006},
journals~\cite{Braun:2006,Bollen:2009} and countries~\cite{Csajbok:2007}
obtained from database of citations. In this context, the \h{} is the
largest integer $h$ such that a node from a given network has at least
$h$ neighbors which have a degree of at least $h$~\cite{Hirsch:2005}.

Korn \etal~\cite{Korn:2009} have proposed a general index to network
node centrality based on the \h{}. Korn~\etal\ named it as lobby index
(\l). Korn~\etal\ argue that the proposed index contains a mix of
properties of other well known centrality measures.  However, they
have studied it mainly in the context of artificial networks like the
Barabasi-Albert model \cite{Barabasi:1999b}.

Like \l{}, degree $D$ is a local centrality measure that is equal to
the number of links of a given node. If the network is directed, the
number of outlinks is the outdegree and the number of inlinks is the
indegree.  Unlike \l{}, betweenness and Eigenvector are global
centrality measures that take into account all nodes in the
network. The betweenness $B$ of a given node is proportional to the
number of geodesic paths (minimal paths between node pairs in the
network) that pass through it.  It seems to be an important measure
for networks where such minimal paths represent transport channels for
information (internet, social networks), energy (power grids),
materials (airports network) or diseases (social and sexual
networks). Eigenvector centrality of a node is proportional to
the sum of the centralities of the nodes to which it is connected,
$\alpha$ is the largest eigenvalue of $A=a_{ij}$ and $n$ the number of
nodes~\cite{bonacich2007}:

\begin{equation}
Ax=\alpha x, \hspace{.5cm} \alpha x_i = \sum^n_
{j=1}a_{ij}x_j, \hspace{.5cm} i = 1, \ldots,n.
\end{equation}

In this paper, we compare the \l{} with degree, betweenness and
Eigenvector centralities applied to associative (non-transport)
networks to obtain the correlation between these measures.

\section{Methods}
\label{sec:methods}
We calculate the \l{}, degree $D$, betweenness $B$ and Eigenvector $E$
centralities for the nodes in linguistic and biological networks
already considered by the physics community. We also plot the
dispersion of $D$ versus $l$, $B$ versus $l$ and $E$ versus $l$, to
verify the correlation between these measures.

We use the linguistic database \textit{Moby Thesaurus
  II}~\cite{MobyII} composed by 30,260 words, for which some network
properties have been studied~\cite{Motter:2002,Holanda:2004}. We choose
the convention that an outlink goes from a root word to a synonym to
construct the network. As an example, in the entry \smallskip

\hfil{\tt set,assign,assign to,assigned,...}\hfill

\smallskip

\noindent the word ``set'' is the root and the link goes to its synonyms. We
obtain the directed links ``set''$\rightarrow$ ``assign'',
``set''$\rightarrow$``assign to'' and ``set''$\rightarrow$ ``assigned''.

The raw thesaurus presents over 2.5 million links, but there are many
words with only inlinks, that is, they are not root words. We worked
with a filtered version containing about 1.7 million links where only
root words constitute nodes. We choose the outlinks to calculate the
centrality measures, and the minimal number of outlinks is 17 and the
maximum is 1,106.

The biological network is the yeast protein-protein network downloaded
from the BioGRID repository~\cite{Biogrid} that is a curated
repository for 5,433 proteins and over 150,000 physical and genetic
unambiguous interactions.

The BioGRID network is composed by gene products connected by a
link~\cite{Biogrid}. The links include direct physical binding of two
proteins, co-existence in a stable complex or genetic interaction as
given by one or several experiments described in the literature. As an
example, using the entries

\smallskip
\begin{tt}
  \centering
  YFL039C YBR243C\\
  YFL039C YKL052C\\
\end{tt}
 \smallskip

 \noindent extracted from BioGRID data set, two links are
 created: \hbox{``YFL039C'' $-$ ``YBR243C''} and ``YFL039C'' $-$
 ``YKL052C'', and the network is undirected.

\section{Results}

\subsection{Local measure: degree}

\begin{figure}[ht]
 \centering
\includegraphics[width=0.8\columnwidth]{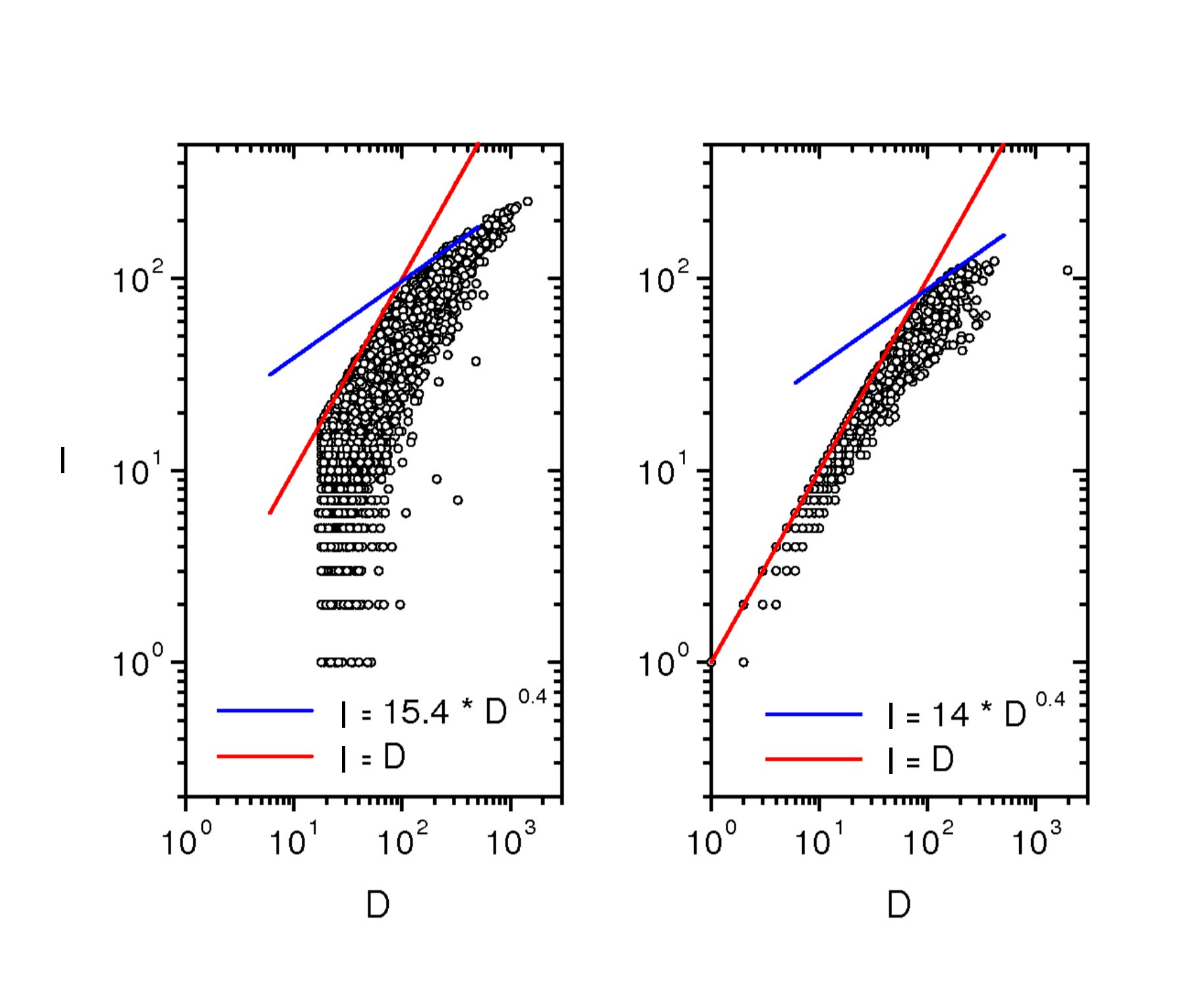}
 \caption{Log-log dispersion plot of $l$ versus Degree centrality $D$
   for a) Moby Thesaurus II and b) Yeast network.
   \label{F1}}
 \end{figure}

In Figure~\ref{F1}, we present dispersion plots of the $l$ versus $D$
for the networks studied.  The \l{} is correlated with $D$ ($h \propto
D$) in the low $D$ regime ($D \leq 100$) in both networks. However,
for higher $D$, one observe $l$ proportional to $D^{0.4}$ for both
networks. The origin of this anomalous exponent is not
clear. Notwithstanding, although correlated, the two measures are not
redundant. In the thesaurus case, the words with low frequency of use
or non-polysemous present low $l$ but high degree.

\subsection{Global measures: betweenness and Eigenvector}

\begin{figure}[ht]
\centering
 \includegraphics[width=0.8\columnwidth]{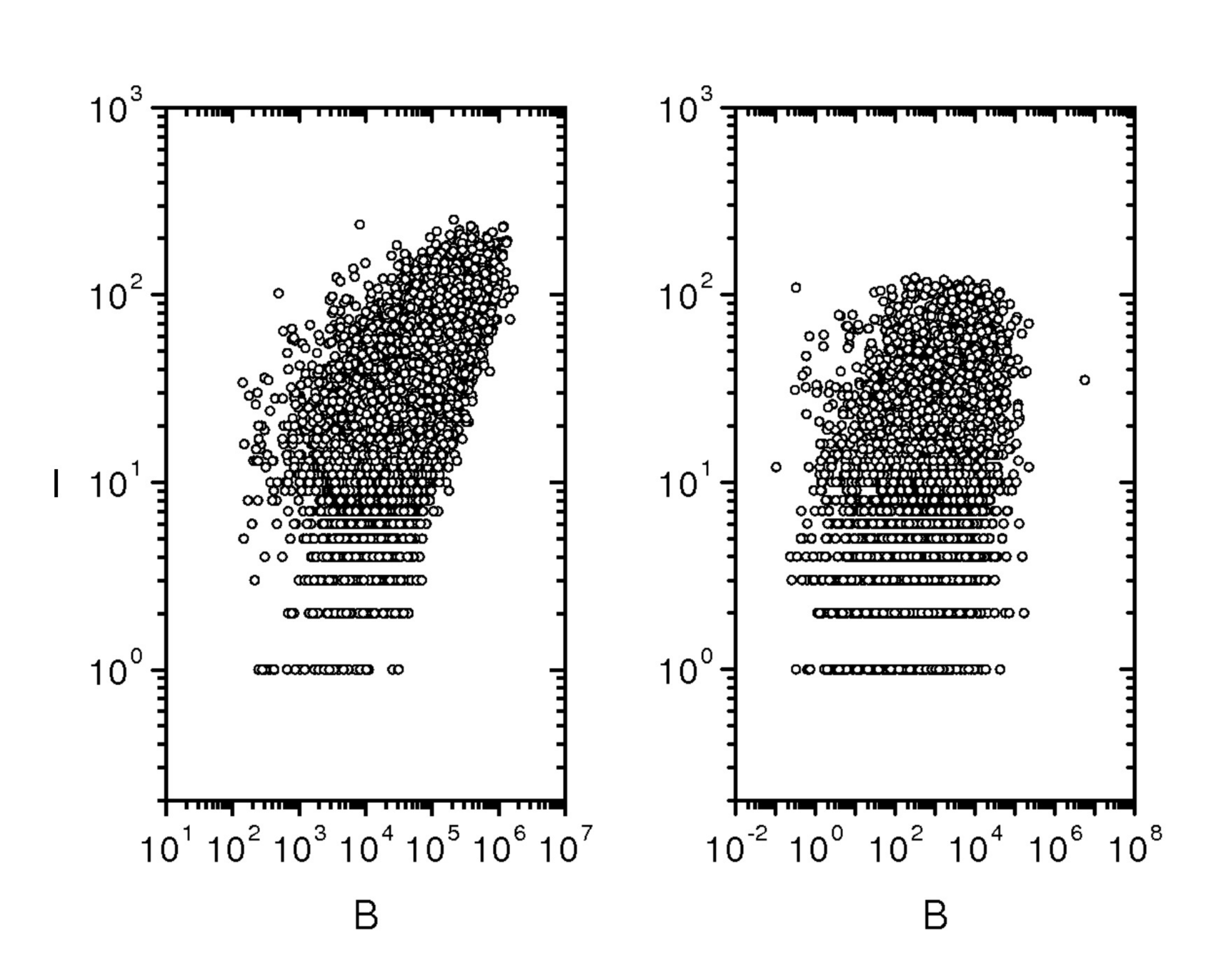}
 \caption{Log-log dispersion plot of $l$ versus Betweenness $B$ for
a) Moby Thesaurus II network and b) Yeast network.}
   \label{F2}
 \end{figure}

We now compare the \l{} with two standard global centrality measures,
betweenness and Eigenvector. First, in Figure~\ref{F2} we present the
dispersion plots of $l$ versus $B$.  The \l{} presents no strong
correlation with $B$ in both networks.

\begin{figure}[h]
 \centering
 \includegraphics[scale=.75]{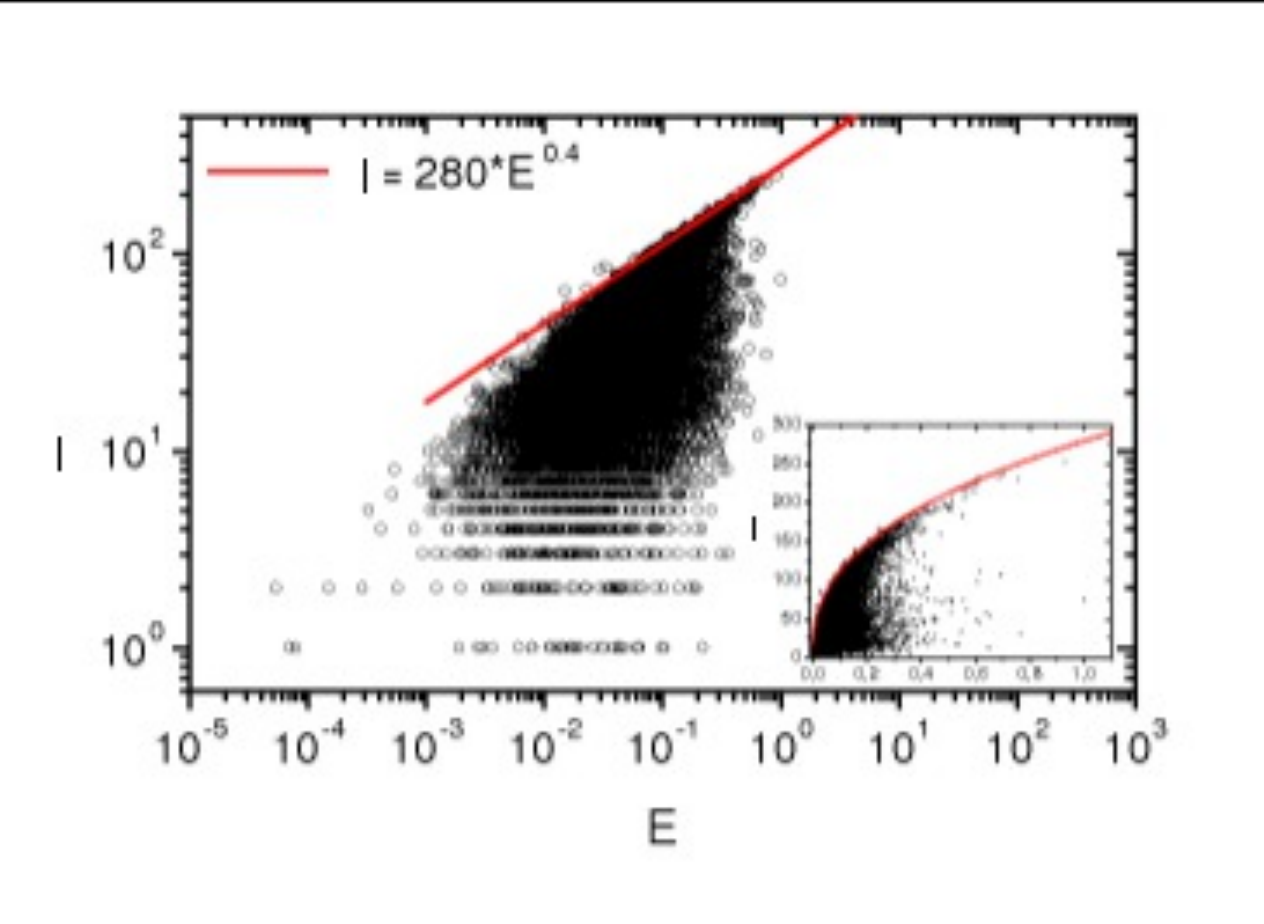}
 \caption{Log-log dispersion plot $l$ versus Eigenvector centrality $E$
for the Moby Thesaurus II. Inset: Linear scale,
notice the several words with high E but low h.} 
\label{F3}
 \end{figure}

 In Figure~\ref{F3}, we give the dispersion plot for the \l{} versus
 the Eigenvector centrality $E$ for the thesaurus network.  In the
 high $E$ regime the maximal $l$ values is bounded by $h \propto
 E^{0.4}$, as in the $l$ versus $D$ plot.  We observe several nodes
 with high $E$ but relatively low $l$ (see Inset).  Examining these
 nodes individually, we find that $l$ seems to outperform $E$ in the
 ranking task, since words with high $l$ also have high $E$ and are
 basic and important polysemous words.  In contrast, terms with high
 $E$ can have high or low $l$. Those with low $l$ are mostly phrasal
 verbs or multiple word expressions derived from the words with high
 $l$.

 It is difficult to qualify a ranking list, but the above effect is
 very clear, as can be observed in Table~\ref{T1}~(see Appendix) that
 shows the top $25$ words ranked by $l$ and $E$, and the same occurs for
 other high $E$ and low $l$ words.

\begin{figure}[ht]
\centering
 \includegraphics[scale=.2]{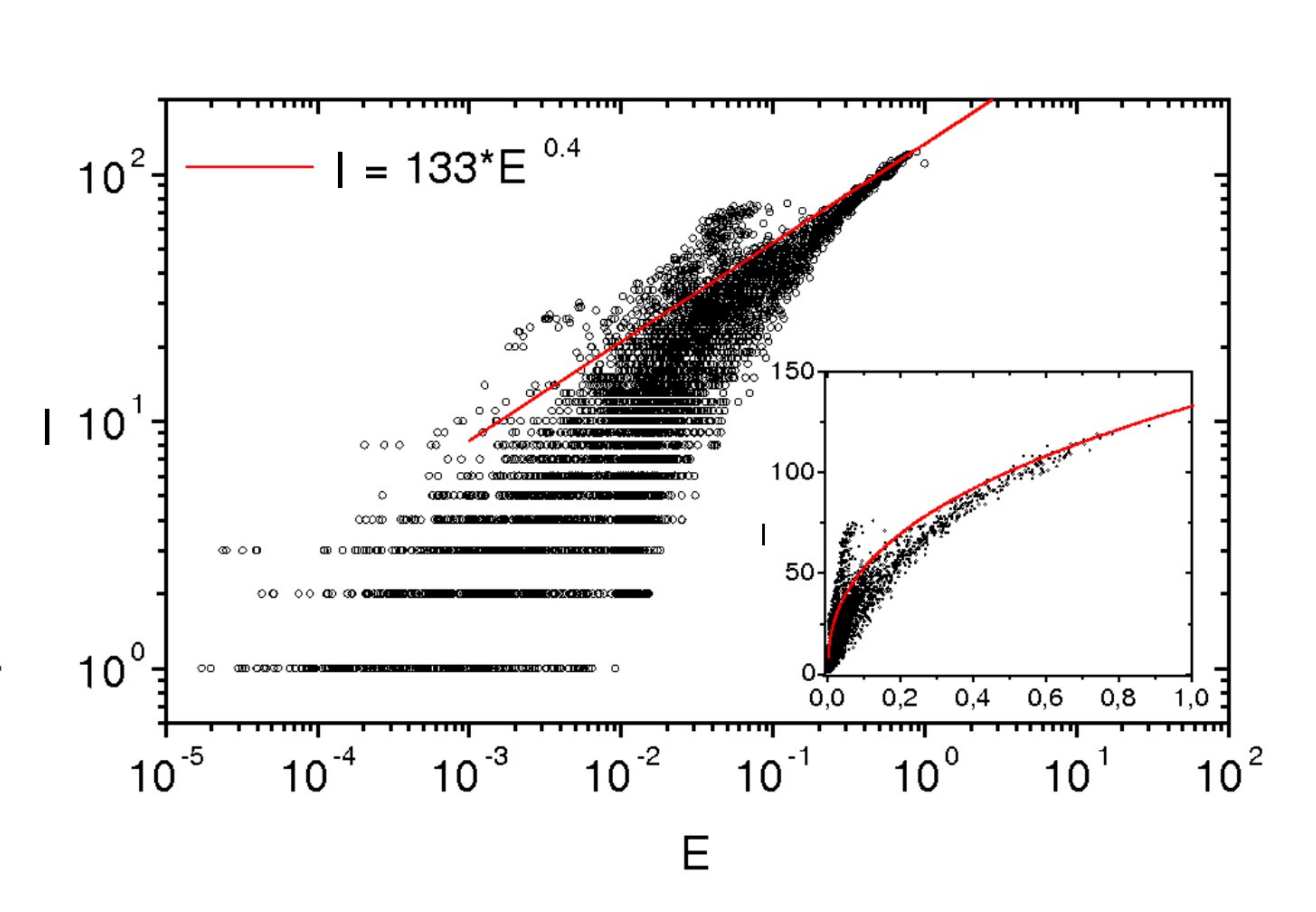}
\caption{Log-log dispersion plot of $l$ versus $E$ for the Yeast network.
 The $l$ and $E$ centralities are well correlated for $E > 0.2$ where there is a
 $h \propto E^{0.4}$ bound for the highest $l$ values. Inset:
linear scale, notice the cluster of high $l$ but low $E$ ribosome proteins.}
\label{F4}
\end{figure}

In the case of the Yeast protein network, we observe a strong
correlation between $l$ and $E$ for $E > 0.2$. The highest $l$ seem
also to be bounded by a $h\propto E^{0.4}$ behavior. Also, the results
suggest that the \l{} could outperform $E$ in the task of classifying
relevant nodes. In the same Figure, one can observe a detaching
cluster of nodes with low $E$ and moderate $l$. We investigated these
nodes and, to our surprise, they all seem to be related to ribosome
assembly, meaning that, somehow, the \l{} carries information that
could be useful in the detection of modules of functionally related
proteins.

\section{Discussion}

In the regime relevant for ranking purposes, the biological network
data shows a strong correlation between the Eigenvector and lobby
centralities, although the computation of the lobby index is much less
demanding because it is not iterative and uses only local
information. This suggests that the $l$ centrality can be useful for
ranking purposes in large databases with results comparable with
Eigenvector centrality.  This claim could be tested in the paper
citation network studied by Chen~\etal~\cite{Chen:2007} where the
Page-Rank algorithm, which core is the Eigenvector centrality,
has given interesting results.

Local measures, such as \l{}, seem to make more sense for
non-transport networks where path distance or channel flux has little
influence and are not important aspects to define
centrality~\cite{Borgatti:2005}.  The same does not occur with some
global measures where path distance must be taken into account. Being
local, \l{} requires $O(D)$ time to compute which is always less than
the $O(NL)$ required to calculate B using Brandes'
algorithm~\cite{brandes2001}, where $N$ is the number of nodes and $L$
is the number of links of a given network. As \l{} requires less
computational time than $E$ ($O(N)$), the high correlation between the
two measures showed for the highest ranks suggests that the \l{} could
be very suitable for ranking tools and search engines.

Both centrality measures make sense for studying diffusion and
epidemic processes in transport networks, but the relevance of minimal
paths is not so clear for linguistic or cultural networks like
thesauri or, as another example, the network of cultural culinary
recipes studied by Kinouchi~\etal~\cite{Kinouchi:2008} where links of
ingredients represent associations but not channels.  For networks
similar to the linguistic one studied here, there is a strong decay of
correlations: two words $A$ and $C$ with minimal path of two links
(that is, $A-B-C$) are almost uncorrelated, since this means that $C$
is not a word semantically related to $A$.  The paths between words
may be relevant to describe perhaps associative psychological
processes (say, $A$ remembers $B$ that remembers $C$), but they are
not channels in the same sense of physical transport networks. So, the
locality of the \l{} could be an advantage to its application for
ranking nodes in non-transport networks where path distance or channel
flux has poor relevance and are not important aspects to define
centrality~\cite{Borgatti:2005}.  We notice that this could be the
case of web pages since links represent more associations than
channels and users do not navigate from link to link by large
distances.

\section{Conclusions}

In conclusion, we studied the \l{} in the Moby II Thesaurus and the
protein-protein interaction Yeast networks.  Several characteristics
of this centrality index have been highlighted. The \l{} seems to be a
better local measure than the node degree $D$ because it incorporates
information about the importance of the node neighbors. Being local,
\l{} requires $O(D)$ time to compute that is always less than $O(N)$
required to compute $E$ and $O(NL)$ time to compute $B$.

We also found that the \l{} is more correlated to Eigenvector
centrality than Betweenness centrality. Indeed, in the ranking task
for words in the thesaurus, \l{} seems even to outperform the $E$ as a
centrality index, detecting basic polysemous words instead of words
with low frequency of use or non-polysemous.
  
Since Eigenvector centrality corresponds to the core idea behind the
original Page-Rank algorithm~\cite{Chen:2007}, which is computationally
very demanding, we suggest that the \l{} could furnish auxiliary
information for ranking pages in the area of Search Engine
Optimization.  Due to the fact that \l{} requires less time to compute
when compared with standard global centrality measures, its use in
other physical, biological and social networks promises very
interesting results.

\bibliographystyle{elsarticle-num}


\section*{Acknowledgements.}
This work received support from CNPq, FAPESP and CNAIPS/US.  The
authors thank to N.~Caticha, R.~Vicente, A.S.~Martinez, R.~Rotta,
P.D.~Batista for suggestions. O. K. thanks his ex-wife C.S.~Motta
for contributing with the original idea for this paper.

\pagebreak
\section*{Appendix}
\begin{table}[h]
  \begin{center}
    \begin{footnotesize}
    \begin{tabular}{c|c|c|c|c|c|c}\hline
      \textbf{$l$ rank} & && & \textbf{$E$ rank} & & \\ \hline
      \textbf{$l$} & \textbf{Eigenvector} & \textbf{Word} & $\:\:\:$& \textbf{Eigenvector} & \textbf{$l$}
      & \textbf{Word} \\ \hline
      252 & 0,930   &  cut    & & 1,000 & 74 & cut up\\
      237 & 0,701     &  set    & & 0,930 & 252 & cut\\
        233 & 0,608     &  run    & & 0,765 & 31 & set upon\\
        232 & 0,687     &  line   & & 0,760 & 230 &turn\\
        230 & 0,760     &  turn   & & 0,701 & 237 & set\\
        225 & 0,598     &  point  & & 0,690 & 106 & break up\\
        222 & 0,608     &  cast   & & 0,687 & 232 & line\\
        220 & 0,584     &  break  & & 0,656 & 54 & line up\\
        218 & 0,560     &  mark   & & 0,649 & 12 & run wild\\
        216 & 0,558     &  measure& & 0,637 & 57 & turn upside down\\
        213     & 0,597 &  pass   & &   0,618 & 112& make up \\
        211 & 0,570 &  check  & &   0,617 & 45 & cast up \\
        209 & 0,487 &  crack  & &   0,608 & 222& cast \\
        206 & 0,562 &  make   & &   0,608 & 233& run \\
        203   & 0,448 &  dash  & &   0,608 &  97& crack up \\
        203 & 0,517 &  stamp   & &   0,604 &  48& check out \\
        202 & 0,514 &  work   & &   0,598 & 225& point \\
        200 & 0,484 &  strain & &   0,597 & 213& pass \\
        196 & 0,491 &  hold   & &   0,584 & 220& break \\
        195 & 0,508 &  form   & &   0,571 &  61& pass up \\
        194 & 0,447 &  beat   & &   0,570 & 211& check \\
        193 & 0,500 &  get    & &   0,562 & 206& make \\
        193 & 0,429 &  rank  & &   0,560 & 218& mark \\
        193 & 0,469 &  round   & &   0,558 &  73& fix up \\
        192 & 0,517 &  go     & &   0,558 & 216& measure \\
        \hline
      \end{tabular}
    \end{footnotesize}
  \end{center}
    \caption{\label{T1} Top 25 words ranked by lobby ($l$) centrality (left) and by
      Eigenvector centrality (right).}
\end{table}

\end{document}